\documentclass[prb,aps,twocolumn]{revtex4}
\usepackage{epsfig}
\usepackage{amsmath}
\draft
\begin{document}
\title{
The stability of a crystal with diamond structure for patchy particles with 
tetrahedral symmetry}

\author{Eva G. Noya}
\affiliation{Instituto de Qu\'{\i}mica F\'{\i}sica Rocasolano, Consejo Superior
	de Investigaciones Cient\'{\i}ficas, CSIC, Calle Serrano 119, 28026 Madrid, Spain}
\author{Carlos Vega} 
\affiliation{Departamento de Qu\'{\i}mica-F\'{\i}sica, Facultad de
	Ciencias Qu\'{\i}micas, Universidad Complutense de Madrid,
	E-28040 Madrid, Spain}
\author{Jonathan P. K. Doye} 
\affiliation{Physical \& Theoretical Chemistry Laboratory, 
             Department of Chemistry, University of Oxford, 
             South Parks Road, Oxford, OX1 3QZ, United Kingdom}
\author{Ard A. Louis} 
\affiliation{Rudolf Peierls Centre for Theoretical Physics, 
             University of Oxford, 1 Keble Road, Oxford, OX1 3NP, 
             United Kingdom}
\date{\today}

\begin{abstract}

The phase diagram of model anisotropic particles with four attractive patches in a tetrahedral arrangement has been computed at two different values for the range of the potential, with the aim of investigating the conditions under which a diamond crystal can be formed.  
We find that the diamond phase is never stable for our longer-ranged potential. At low temperatures and pressures, the fluid freezes into a body-centred-cubic solid that can be viewed as two interpenetrating diamond lattices with a weak interaction between the two sublattices. 
Upon compression, an orientationally ordered face-centred-cubic crystal becomes more stable than the body-centred-cubic crystal, and at higher temperatures a plastic face-centered-cubic phase is stabilized by the increased entropy due to orientational disorder.   
A similar phase diagram is found for the shorter-ranged potential, but at low temperatures and pressures, we also find a region over which the diamond phase is thermodynamically favored over the body-centred-cubic phase.  
The higher vibrational entropy of the diamond structure with respect to the body-centred-cubic solid explains why it is stable even though the enthalpy of the latter phase is lower. 
Some preliminary studies on the growth of the diamond structure starting from a crystal seed were performed.  
Even though the diamond phase is never thermodynamically stable for the longer-ranged model, direct coexistence simulations of the interface between the fluid and the body-centred-cubic crystal and between the fluid and the diamond crystal show that, at  sufficiently low pressures, it is quite probable that in both cases the solid grows into a diamond crystal, albeit involving some defects.   
These results highlight the importance of kinetic effects in the formation of diamond crystals in systems of patchy particles.
\end{abstract}

%
\maketitle

\vspace{0.5cm}
\section{Introduction}

	The study of the behaviour of anisotropic particles has
attracted significant attention in recent years. 
Initially the interest arose because they were seen as very simplified models
of proteins.\cite{sear_jcp1999,song_pre2002,dixit_jcp2002,JCP_118_9882_2003,sandler_jcp2004,talanquer_jcp2005} 
Even though important advances have been made using simple isotropic models,\cite{rosenbaum,tenwolde}
interactions between proteins are highly anisotropic,\cite{Matthews68} and
an improved description of the behaviour of proteins can potentially be obtained
using models that explicitly incorporate this 
anisotropy.\cite{sandler_jcp2004,jon,JCP_127_084902_2007,JCP_130_084902_2009,PNAS_104_16856_2007,JCP_129_085102_2008,JCP_131_115101_2009}
For example,  anisotropic models can lead to the stabilization of low density
crystals,\cite{sandler_jcp2004,jon}
with packing fractions similar to those typically formed by 
proteins.\cite{Matthews68}
They are also able to reproduce quantitatively the metastable 
fluid-fluid phase separation of globular
proteins,\cite{JCP_127_084902_2007,JCP_130_084902_2009} 
whereas isotropic models could only reproduce it qualitatively.
In addition it has been suggested that kinetics of 
protein crystallization could be sensitive to the degree of anisotropy.\cite{JACS_122_156_2000}

	In the last few years, a number of experimental groups have developed
new methods to produce nanoparticles and colloids with anisotropic shapes or 
interactions,\cite{vanderhoff,pine_sicence2003,Jackson04,pine_jacs2005,pine_chemmat2005,Roh05,Snyder05,Li05b,blaaderen_science2006,Roh06,Devries07,Cho07,langmuir_24_621_2008,Perro09,SM_5_3823_2009} 
and this has led to increased interest in anisotropic particles.\cite{nm_6_557_2005,Yang08,Pawar10}  
These experimental developments have motivated many theoretical and
simulation studies on how these patchy particles would assemble into
crystalline structures\cite{sandler_jcp2004,jon,JCP_127_054501_2007,JPCB_113_15133_2009,glotzer_langmuir2005,Doppelbauer10} or 
into clusters with a particular geometry.\cite{chandler,JCP_127_085106_2007,JCP_131_175101_2009,JCP_131_175102_2009}
Much of the work in this latter topic has been also aimed at
getting a better understanding of the assembly of virus capsids.\cite{chandler,JCP_131_175102_2009}
One interesting example of the degree of complex behaviour that anisotropic
particles can exhibit is provided by one patch particles that mimic
Janus particles (colloidal
particles whose surface is divided in two areas with
different chemical composition)\cite{JCP_131_174114_2009,PRL_103_237801_2009} 
which have been shown to exhibit simultaneously gas-liquid phase
separation and the formation of micelles.\cite{PRL_103_237801_2009}
This recent example illustrates the potential richness of the behaviour that
anisotropic models can exhibit, and that there is much still to be learnt.

	In previous work, we have studied the crystallization behaviour of
patchy particles in two and three dimensions and found that 
the geometry of the patches strongly affects
crystallization.\cite{jon}  
For example, crystallization can be frustrated when the patches are not 
straightforwardly compatible with a crystalline structure, 
e.g.\ five regularly-arranged patches for two dimensional particles. 
However, perhaps more surprising is that even in cases where the symmetry of 
the particles is compatible with a crystalline structure, there can be strong 
variations in the crystallization behaviour.
In particular, we found that whereas a simple-cubic structure can be
easily obtained by quenching a fluid of six-patch octahedral particles, 
it is difficult to obtain a diamond structure by quenching a fluid of four-patch tetrahedral particles.\cite{jon}
Similarly, Zhang \emph{et al.} were only able to obtain a diamond
crystal from such tetrahedral patchy particles when a crystal seed was 
inserted in the simulation box or when the
model potential included torsional interactions.\cite{glotzer_langmuir2005}
By studying the geometry of the clusters formed by the octahedral and tetrahedral
model particles, Doye \emph{et al.} attributed the different behaviour of the two systems to
the frustration between the local order in the fluid and the global crystalline
order for the tetrahedral particles.\cite{jon}
Given the possible applications of a diamond colloidal crystal in photonics
due to its predicted optical band gap,\cite{Maldovan03,Maldovan04,Hynninen07} and the
growing interest in patchy particles in general, it would be interesting to 
study in more detail the crystallization behaviour of the tetrahedral patchy
particles.

In this work the phase diagram of model tetrahedral particles is investigated
by means of computer simulation. Even though there is a very recent study
on the phase behaviour of tetrahedral patchy particles,\cite{JPCB_113_15133_2009} 
the present work differs in the model used to describe patchy particles. 
Romano \emph{et al.}\cite{JPCB_113_15133_2009} used the Kern-Frenkel (KF)
model,\cite{JCP_118_9882_2003} in which particles
are described as hard spheres with some attractive sites modeled as square 
wells, whereas in this work particles are modeled using a generalized 
Lennard-Jones
potential modulated by Gaussian functions at the location of the patches.\cite{jon}
This model potential (and modified versions of it) has been previously used 
to study crystallization,\cite{jon} phase behaviour\cite{JCP_127_054501_2007} 
and the self-assembly of clusters of patchy
particles with various symmetries.\cite{JCP_127_085106_2007,JCP_131_175101_2009,JCP_131_175102_2009}
Comparisons between the present work and that of Romano \emph{et al.} will
allow us to discern the intrinsic behaviour of tetrahedral particles 
from particular behaviour that arises from the specific shape of the model
potential.  

\section{Method}

\subsection{Model}

	Anisotropic particles are modeled using a pair potential 
that consists of a generalized Lennard-Jones (LJ) repulsive core and an attractive tail
modulated by an angular function that depends on how directly the patches 
point at each other. 
The interaction between two particles $i$ and $j$ depends on the distance
vector between them (${\bf r}_{ij}$) and on their orientation
(${\bf \Omega}_i$ and ${\bf \Omega}_j$):
\begin{equation}
V({\bf r}_{ij},{\bf \Omega}_i,{\bf \Omega}_j)  =
\begin{cases}
V_{\rm LJ}(r_{ij})  & r_{ij} <  \sigma_{\rm LJ} \\
V_{\rm LJ}(r_{ij})V_{\rm ang}(\widehat{\mathbf{r}}_{ij},{\bf \Omega}_i,{\bf \Omega}_j) & r_{ij} \ge  \sigma_{\rm LJ}
\end{cases}
\label{eq2}
\end{equation}
where $V_{\rm LJ} (r_{ij}) $ is a generalized $2n-n$ LJ potential:
\begin{equation}
V_{\rm LJ}(r_{ij}) = 4 \epsilon \left[ \left( \frac{\sigma_{\rm LJ}}{r_{ij}} \right)^{2n}
		-\left( \frac{\sigma_{\rm LJ}}{r_{ij}} \right)^{n} \right],
\end{equation}
and $\sigma_{\rm LJ}$ is the distance at which the LJ potential passes through zero.
Our purpose is to study the phase behaviour for the usual 12-6 LJ model. 
However, we are also interested in investigating the effects of the range of 
the potential, which can be tuned by modifying the exponents of the 
generalized LJ potential. 
Although the depth of the potential is independent of the
value of $n$, the position of the potential minimum varies and is at 
$2^{1/n}\sigma_{\rm LJ}$.
The phase behaviour of the 20-10 model will also be investigated in this 
work. For this model, the position of the minimum is
1.0718$\,\sigma_{\rm LJ}$, whereas for the usual 12-6 LJ potential it is 
1.1225$\,\sigma_{\rm LJ}$.

The generalized LJ potential is modulated by the factor 
$V_{\rm ang} (\widehat{\mathbf{r}}_{ij},{\bf \Omega}_i,{\bf \Omega}_j)$, 
which is a product of Gaussian functions that depends on the allignment of the 
patches with the interparticle vector:
\begin{equation}
V_{\rm ang} (\widehat{\mathbf{r}}_{ij},{\bf \Omega}_i,{\bf \Omega}_j) = 
\exp \left( - \frac{\theta_{ijk_{\rm min}}^2}{2\sigma_{\rm pw}^2} \right)
          \exp \left( - \frac{\theta_{jil_{\rm min}}^2}{2\sigma_{\rm pw}^2} \right)
\end{equation}
where $\theta_{ijk}$ is the angle between $\widehat{\mathbf{r}}_{ij}$ and
patch $k$ on particle $i$, and $k_{\rm min}$ is the patch that minimizes this
angle.
Thus, $V_{\rm ang}=1$ when two patches directly point at each other.
The parameter $\sigma_{\rm pw}$ is a measure of the width of the patches, 
with $2\sqrt{2}\,\sigma_{\rm pw}$ being the full width at half maximum of the Gaussian.
For computational efficiency, this potential was truncated and shifted 
at a cutoff distance of $2.5\,\sigma_{\rm LJ}$.

In this work, we study particles with four tetrahedrally-arranged patches 
with a patch width of $\sigma_{\rm pw}$=0.3 radians. 
In a previous study of octahedral particles with 6 patches using the 
12-6 model it was found that this patch width is sufficiently narrow 
to stabilize a low density simple-cubic crystal.\cite{JCP_127_054501_2007}
Therefore, we expect these particles to represent promising candidates for 
the formation of a diamond crystal at low pressure.

Throughout this paper, all quantities will be given in reduced 
units, i.e.\ $u^*=u/\epsilon$, $T^*=k_B T/\epsilon$,  $\rho^*=\rho\sigma_{\rm LJ}^3$,
and $p^*=p\sigma_{\rm LJ}^3/\epsilon$.

\begin{figure}[!tbhh]
\begin{center}
\includegraphics[width=85mm,angle=0]{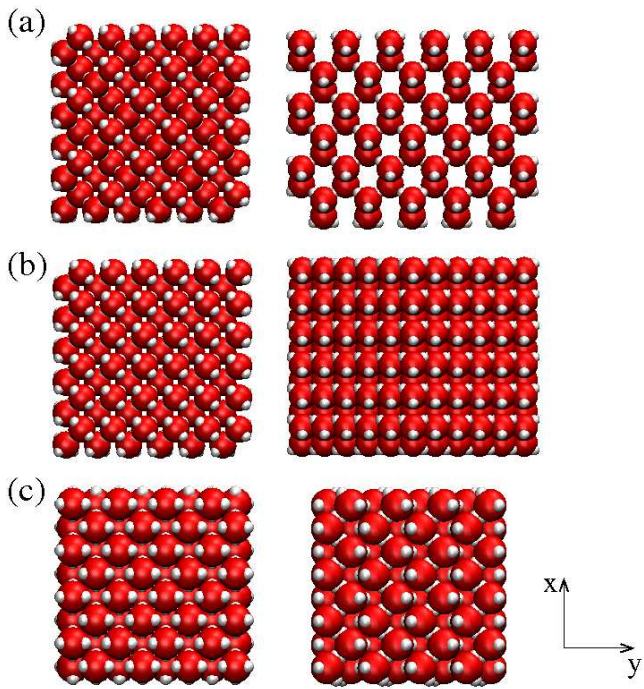}
\caption{\label{figure_solids} Orientationally-ordered crystal structures 
for tetrahedral patchy particles: (a) diamond, (b) bcc and (c) fcc. Two views
are shown in each case, with the picture on the right corresponding to a 
rotation of the structure by (a) and (b) $\pi/4$ and (c) $\pi/2$ about
the $x$ axis.}
\end{center}
\end{figure}

\subsection{Solid structures}
\label{subsect:solids}

	As mentioned before, the tetrahedral geometry of the particles was 
chosen in order to explore the possibility 
that, at sufficiently low temperatures and pressures, the formation of
a diamond crystal would be favoured. In the diamond lattice
each of the four patches point directly at one of the four nearest
neighbours (Fig.\ \ref{figure_solids}(a)). 
As the four patches are able to form a ``perfect'' bond (i.e.\
with energy $-\epsilon$), the energy will be minimized in this structure.

However, the diamond solid has a very low density and 
it is expected that a body-centred-cubic (bcc) structure
(Fig.\ \ref{figure_solids}(b)) will be competitive with diamond.  
Specifically, in the diamond  crystal there is enough free space to 
interpenetrate a second diamond lattice displaced with respect to the first 
one by a vector $(a/2,a/2,a/2)$, $a$ being the unit cell parameter of the 
cubic diamond lattice, and the density of the resulting bcc crystal is 
exactly twice that of the diamond structure. 
When the distance between nearest neighbours is equal to $\sigma_{\rm LJ}$, 
i.e.\ when the repulsive cores of the particles touch, then
the density of the diamond and bcc crystals are $\rho^*$=0.6495 and 1.2990,
respectively. However, the energetically-preferred nearest-neighbour distance
corresponds to the minimum in the potential, and the corresponding
density of the crystals are $\rho^*$=0.4593 and 0.9186 for the 12-6 model 
and $\rho^*$=0.5276 and 1.0551 for the 20-10 model.
Consequently, for both the models that we consider,
it is always possible to interpenetrate the two sublattices without
any deformation of the two lattices, i.e.\ with no energy cost,
although there is slightly less room for the sublattices to vibrate in the 
shorter-ranged model.
This situation is somewhat similar to that found for the octahedral particles,
in which case the bcc structure is formed by interpenetrating two simple-cubic
lattices.\cite{JCP_127_054501_2007} 
However, for the octahedral particles the two simple cubic lattices have to be
expanded slightly with respect to their ideal densities in order to 
interpenetrate without the repulsive cores of the particles overlapping, and
so there is an associated energy penalty.\cite{JCP_127_054501_2007}

At zero temperature and zero pressure the most stable solid will be
that with least potential energy. For the patch width studied in
this work there is a small attractive interaction between the
two diamond sublattices in the bcc structure which slightly lowers 
the energy of the bcc solid with respect to that of the diamond.
Therefore, at zero temperature and at zero pressure the bcc solid
is the most stable solid. This represents a difference with
the Kern-Frenkel\cite{JCP_118_9882_2003} 
model studied by Romano \emph{et al.}\cite{JPCB_113_15133_2009}, for
which the bcc and diamond solids exhibit the same
energy (both maximize the number of bonds per
particle) and so they are degenerate at zero temperature and
at zero pressure.
At zero temperature and pressures above
zero, the more stable phase will be that with lower enthalpy. As
the molar volume of the bcc solid is lower than that for 
diamond, the $pV$ term is lower for the bcc solid and, therefore,
the bcc solid is again more stable than the diamond structure. In summary,
for our model, at zero temperature the diamond structure becomes
more stable than the bcc solid only at negative pressures. 
At zero pressure and finite temperatures, however, it is probable that
the diamond structure has a higher vibrational entropy, because interactions
between the two sublattices are likely to  
reduce the vibrational entropy of the bcc solid, 
i.e.\ the atoms in the bcc solid have less ``room'' to vibrate because of 
the presence of the other sub-lattice.
Therefore, it is possible that the diamond structure could be 
stabilized if the entropy term, which is expected to be somewhat higher in
the diamond structure, overcomes the advantage in the potential energy
of the bcc solid.

For completeness the high pressure region of the phase diagram will also be 
studied.
Similar to what has been found for octahedral 
particles,\cite{JCP_127_054501_2007}
it is expected that at high pressures a face-centred-cubic (fcc) solid
is the most stable phase. For tetrahedral particles it is not possible
to align each of the four patches with a nearest neighbour. 
However, an ordered fcc structure (fcc-o) can be obtained
by starting from the bcc lattice described above and then stretching one of
the edges of the unit cell from $a$ to $a\times\sqrt{2}$ so that each one of the
four patches will be pointing to four of the twelve nearest neighbours,
but the allignment will not be perfect (Fig.\ \ref{figure_solids}(c)). 
This structure has a somewhat higher energy
than the diamond or the bcc lattices. It is expected to become the stable 
phase above some given pressure, where this high density structure
will be favoured by its lower enthalpy (the $pV$ term will compensate
for the disadvantage in the potential energy). 

At high temperatures, where the kinetic energy is high enough to overcome
the attractive interactions, it is likely that the ordered fcc will transform
into a plastic crystal in which the centers of mass of the particles are 
arranged in an fcc lattice but where the particles are free to rotate.
The plastic crystal will be denoted fcc-d.

\subsection{Details of the simulations}
	
$NpT$ Monte Carlo (MC) simulations were used for both the fluid and the solid phases. Typically about 200\,000 MC cycles (plus
another 100\,000 MC cycles for equilibration) were used for the fluid phase, 
whereas 50\,000 MC cycles (plus 50\,000 MC cycles for equilibration) were 
enough for the solid phases. Each MC cycle
consisted of $N$ attempts to translate or rotate a particle ($N$ being the
number of particles in the system) plus one attempt to change the volume.
The maximum translational and rotational displacements were adjusted
to obtain a 40\% acceptance probability and the maximum volume displacement was
adjusted to obtain a 30\% acceptance probability.
The number of particles used in the simulations was 512 for
the diamond crystal, 432 for the bcc lattice, 500 for the fcc solid
and 432 for the fluid phase. These numbers are chosen so that the crystal structures are commensurate with the simulation boxes.

The computation of the phase diagram requires the calculation of free energies.
As the methods used to compute free energies in this work have 
been previously described in detail,\cite{review,vega_noya,JCP_129_104704_2008}
only a brief summary will be given here. 
For the fluid the free energy was calculated by thermodynamic integration with 
the ideal gas as a reference state.\cite{frenkelbook} 
Between 10--20 states were used in the integration.
For the solid phases we used the recently
proposed Einstein molecule approach,\cite{review,vega_noya,JCP_129_104704_2008} 
which is a variant of the Einstein crystal
method of Frenkel and Ladd.\cite{frenkel-ladd,JCP_112_5339_2000} 
In this method, the free energy of the solid
is calculated by Hamiltonian integration with the reference state being
an Einstein molecule (i.e.\ an Einstein crystal in which one of the molecules, 
e.g.\ molecule 1, does not vibrate) with the same lattice as the real solid.  
As we are considering anisotropic particles,
besides the harmonic springs that bind the center of mass of each particle to
a lattice position, an orientational field that keeps the particles with the 
right orientation is also needed. 
It is convenient to choose an orientational
field with the same symmetry as the model under study.\cite{review,JCP_112_8950_2000} 
The orientation of each particle in the reference structure is defined by two
unitary vectors ${\bf a_0}$ and ${\bf b_0}$ (non-orthogonal) parallel to two specified patches.
For the tetrahedral particles (that exhibit T$_d$ symmetry), the reference system will
be:
\begin{eqnarray}
\label{field_orient}
& & U_{Ein-mol}=U_{trans}+U_{orient} = \\  
& & \sum_{i=2}^N \lambda_t ({\bf r}_i - {\bf r}_{i,0} )^2 + \sum_{i=1}^N \lambda_o \left[ \sin^2 
\left( \Psi_{a,i} \right) + \sin^2 \left(\Psi_{b,i}\right) \right] \nonumber
\end{eqnarray}
where $\lambda_t$ and $\lambda_o$ are the coupling parameters,
${\bf r}_i$ is the instantaneous position of the center of mass of
molecule $i$ and ${\bf r}_{i,0}$ is its equilibrium position.
$\Psi_{a,i}$ is the angle formed by the closest patch in the instantaneous
orientation of molecule $i$ and the vector ${\bf a_0}$ in the reference structure, and
$\Psi_{b,i}$ is defined analogously. Note that the second sum in Eq.\ 
\ref{field_orient} runs over all the particles,
i.e.\ all molecules are allowed to rotate.\cite{review,vega_noya}
The free energy of the reference system and the 
free energy difference between the reference system and the solid was
evaluated by using the procedure described in 
Refs.\ \onlinecite{review} and \onlinecite{JCP_129_104704_2008}.

	Once the free energy is known at a given thermodynamic state, the
free energy can be computed at other states by thermodynamic integration.\cite{frenkelbook,review}
Coexistence points were calculated by imposing the conditions of
chemical equilibrium, i.e.\ equal temperature, pressure and chemical potential.
Starting from the coexistence points calculated
by free energy calculations,
Gibbs-Duhem integration\cite{kofkeJCP,kofkeMP} with a fourth order Runge-Kutta algorithm\cite{recipes} was used 
to trace the coexistence lines.


	The melting point 
of the diamond and bcc solids was also calculated by using the direct coexistence 
method.\cite{woodcock1,woodcock2,JCP_127_054501_2007} 
We follow the same procedure as that described in Ref.\ \onlinecite{JCP_127_054501_2007}.
In this method, simulations of a fluid-solid interface are performed.
For the fluid-diamond interface, the initial configuration 
contained 512 solid particles (i.e.\ 4$\times$4$\times$4 unit cells)
and 512 fluid particles. The fluid-bcc interface
contained 432 solid particles (i.e.\ 6$\times$6$\times$6 unit cells)
and another 432 fluid particles. Finally, in the fluid-fcc-d interface
there was a crystalline block of 500 particles 
(i.e.\ 5$\times$5$\times$5 unit cells)
plus another 500 fluid particles. 
The interfaces were generated as in Ref.\ \onlinecite{JCP_127_054501_2007}.
The coexistence point was then calculated by performing $NpT$ MC simulations 
at a given temperature and at different pressures. Monitoring the evolution
of the internal energy or the density, it is possible to bracket the coexistence
pressure at the simulated temperature. 
Alternatively, at a given pressure, simulations can be performed
at various temperatures to bracket the coexistence temperature at that pressure.

\section{Results}
\label{results}

Let us start with the results for the long-ranged 12-6 LJ
tetrahedral model.
Before presenting the computed phase diagram,
we first consider the fluid-fluid phase equilibrium. 
In a previous
study of octahedral particles using the same model potential as the
one used here, it was found that for a patch width of $\sigma_{\rm pw}=$ 0.3 radians
fluid-fluid phase separation
was metastable with respect to solidification.\cite{JCP_127_054501_2007}
As shown in previous work,\cite{JCP_118_9882_2003} the fluid-fluid phase separation moves 
to lower temperatures as the surface coverage of the patches
diminishes.  From this result it follows that, for the same patch width,
the fluid-fluid phase separation for
the tetrahedral model (four patches) occurs at a lower temperature than 
that of the octahedral model (six patches). As 
fluid-fluid phase separation was already metastable
for the octahedral model at a patch width of $\sigma_{\rm pw}=$0.3 radians,\cite{JCP_127_054501_2007} 
it is likely that it is also metastable for the tetrahedral model.
For this reason, studies of fluid-fluid phase separation were not
attempted in this work, although it would be interesting to study
the emergence of an equilibrium liquid phase at larger 
$\sigma_{\rm pw}$ in future work.

\begin{table}[t]
\centering
\caption{\label{tbl_free_energies} Helmholtz free energies ($A_{sol}$) 
of the solid phases for the 12-6 model, as obtained by the Einstein molecule method.
The coupling parameters in the Einstein
molecule were chosen as $\lambda_t/(k_BT/\sigma_{\rm LJ}^2)=\lambda_o/(k_BT)$=30000.
The free energy $A_{sol}$  and the 
average potential energy energy ($U$) are given in units of $Nk_BT$. 
The uncertainty in $A_{sol}$ and $U$ is about 0.02$NkT$.
(TI) refers to independent simulations run as consistency checks. 
It can be seen that the free energy obtained by thermodynamic
integration along an isotherm coincides with the value calculated
from the Einstein molecule method within statistical uncertainty.}
\begin{tabular}{lcccccccccc}
\hline\hline
Structure & & $T^*$  & & $p^*$ & & $\rho^* $ & & $U$ & &  A$_{sol}$ \\
\hline
  diamond      & & 0.10  & & 0.3  & & 0.466  & &  -16.30  & & -5.09 \\
  diamond      & & 0.10  & & 1.2  & & 0.513  & &  -14.95  & & -3.69 \\
  diamond (TI) & & 0.10  & & 1.2  & & 0.513  & &           & & -3.71 \\
\hline
  bcc          & & 0.10  & & 3.00 & & 1.035  & &  -15.28  & & -3.34 \\
  bcc          & & 0.10  & & 0.65 & & 0.932  & &  -16.56  & & -5.11 \\
  bcc (TI)     & & 0.10  & & 3.00 & & 1.035  & &           & & -3.35 \\
  bcc          & & 0.15  & & 0.5  & & 0.900  & &  -9.70   & & -0.03  \\
\hline
  fcc-o        & & 0.10  & & 6.08 & & 1.234  & &  -8.31   & & 3.17 \\
\hline
  fcc-d        & & 0.50  & & 10.0 & & 1.208  & &   0.12   & &  5.98 \\
 \hline\hline
\end{tabular} 
\end{table}

	We focus now on the fluid-solid and solid-solid
phase separation. Helmholtz free energies 
were calculated for all the considered solid phases at some specified thermodynamic states
(see Table \ref{tbl_free_energies}).
These free energy calculations were tested against thermodynamic consistency
checks. Once the free energy is known at a particular thermodynamic state, coexistence points
can be obtained by thermodynamic integration (see Table \ref{tbl_CP}).
Let us consider 
the results for $T^*=$0.1. At this temperature there is a
phase transition at almost zero pressure from a very low density fluid to the bcc solid,
which upon increasing the pressure transforms into the
orientationally-ordered fcc-o phase.
The bcc phase is stabilized with respect to the fluid and the fcc-o solid
by a low average potential energy.
The fcc-o solid has an appreciably higher energy than the bcc (see Table \ref{tbl_CP}),
but for sufficiently high pressures the fcc-o becomes more favourable due to
its higher density and entropy (which is probably due to the greater 
orientational freedom that arises because the patches cannot perfectly align 
with all the nearest neighbours).

\begin{figure}[tb]
\begin{center}
\includegraphics[width=85mm,angle=0]{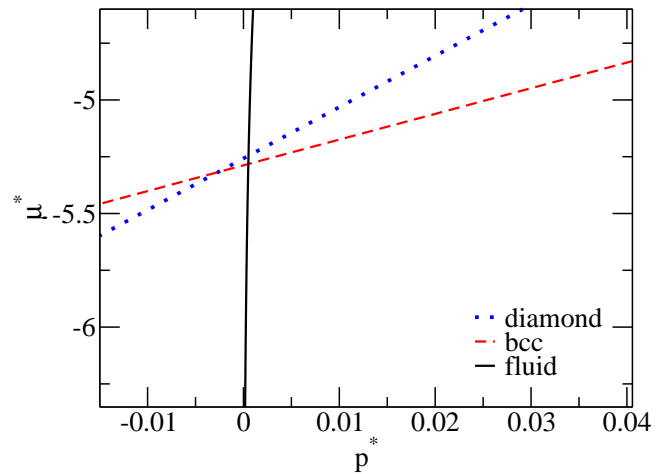}
\caption{\label{mu} Chemical potential ($\mu^*$) as a function
	of pressure for the fluid phase
	and the diamond and bcc solids for the 12-6 LJ model
	along the $T^*=$0.1 isotherm.}
\end{center}
\end{figure}

\begin{table}[!htb]
\centering
\scriptsize
\caption{\label{tbl_CP} Coexistence points for the 12-6 model obtained using 
        thermodynamic integration together with the Helmholtz free 
        energies given in Table \ref{tbl_free_energies}.
	Uncertainties in the potential energy per particle
	$u^*$ are smaller than 0.01.}
\begin{tabular}{llccccccccc}
\hline\hline
     Phase 1  & Phase 2  & $T^*$ & $p^*$ & $\rho^*_1$ & & $u^*_1$ & & $\rho^*_2$ & & $u^*_2$ \\
\hline
     fluid   & bcc    & 0.10 &  0.0005(2) & 0.0005(2) & & -0.07 & & 0.882(1) & & -1.64 \\
     diamond & bcc    & 0.10 &  -0.003(2) & 0.441(1)  & & -1.62 & & 0.882(1) & & -1.64 \\
     bcc     & fcc-o  & 0.10 &  4.06(5)   & 1.068(1)  & & -1.44 & & 1.179(1) & & -1.00 \\
\hline
     fluid   & bcc    & 0.15 &  0.21(5)   & 0.554(1)  & & -0.35 & & 0.873(1) & & -1.43 \\
     fluid   & fcc-d  & 0.50 &  5.91(5)   & 0.985(1)  & & 0.02  & & 1.082(1) & & -0.04 \\
     fluid   & fcc-d  & 1.00 &  13.1(3)   & 1.036(1)  & & 0.58  & & 1.136(1) & & 0.42 \\
 \hline\hline
\end{tabular}
\end{table}

\begin{figure}[tb]
\begin{center}
\includegraphics[width=85mm,angle=0]{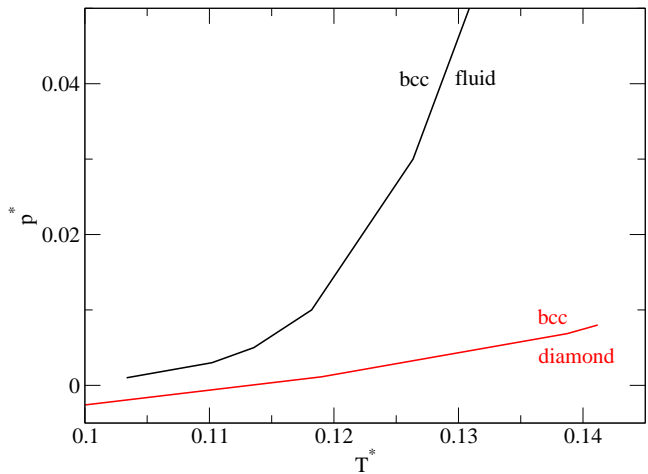}
\caption{\label{diamond_stability} Fluid-bcc and diamond-bcc coexistence lines for
	the 12-6 LJ model obtained
	from Gibbs-Duhem simulations. Above $T^*=$0.14 the diamond solid is
	not mechanically stable.}
\end{center}
\end{figure}

We now consider the stability of the diamond structure at this temperature. 
The chemical potentials for the fluid phase, and the diamond and bcc solids 
along the $T^*=$0.1 isotherm are shown in Fig.\ \ref{mu}. 
Our results predict that the diamond structure only becomes more stable than the
bcc solid at slightly negative pressures 
(see Table \ref{tbl_CP}), 
in a region of the phase diagram where the fluid phase
is more stable than both the diamond and the bcc solids.
We should note that as the difference in pressure between the 
fluid-bcc and the diamond-bcc coexistence points is so small at this 
temperature, the accuracy of the present calculations does not allow us 
to totally rule out the possibility that these transitions occur in a 
different order. 
However, even if this were the case the diamond crystal would at most
only be marginally stable at very low pressures at this temperature.
Moreover, as mentioned in Section \ref{subsect:solids}, the metastability of the
diamond structure is not unexpected,
because the bcc solid shows a slightly lower energy and also a lower
$pV$ term due to its higher density.
Therefore, the diamond structure could only be stabilized if
it would exhibit a higher vibrational entropy than the 
bcc solid to overcome the lower enthalpy of the latter. 
It is reasonable to think that diamond might have a somewhat higher 
entropy because it is likely that in the bcc solid the movement
of the particles is somewhat impeded by interactions 
between the two diamond sublattices. 
If correct, the diamond structure should then
gain stability with respect to the bcc solid as the temperature is increased
and so might even become the thermodynamically stable phase at higher 
temperature.
To check this possibility, the fluid-bcc and
the diamond-bcc coexistence lines were calculated using Gibbs-Duhem
integration. As shown in Fig.\ \ref{diamond_stability}, the diamond
crystal indeed gains some stability with respect to the bcc solid as the
temperature increases, although the effect is relatively small. 
At higher temperatures the diamond/bcc coexistence occurs at 
slightly positive pressures.
However, this happens in a region where the most stable phase is the
fluid, i.e.\ after the sublimation of the bcc solid.

The absence of a region in the phase diagram in which the diamond solid
is the most stable phase differs from recent calculations of the phase diagram of 
tetrahedral particles performed by 
Romano \emph{et al.}\cite{JPCB_113_15133_2009} However,
these authors have used 
a different model potential to describe the interactions
between the tetrahedral particles and most likely the differences between the present
work and that of Romano \emph{et al.} are due to the use of
a different model potential. 
Indeed, Vega and Monson using the primitive model of water (PMW)\cite{kolafa} 
that bears some resemblance to the KF model also found that the diamond 
lattice was thermodynamically stable.\cite{vegamonson} 
The PMW particles are also hard-spheres with four patches in a tetrahedral 
arrangement, whose interactions are modelled using square-well potentials.
However in the PMW there are two inequivalent types of patches, 
and only patches of different type interact.
The possible origin of the differences between our work
and that of Romano \emph{et al.} will be discussed in more
detail later.

\begin{table}[t]
\centering
\caption{\label{tbl_GD} Coexistence points for the 12-6 model obtained using the
        Gibbs-Duhem method. Uncertainties in the densities $\rho^*$ are of the order of
        0.001 and in the potential energy per particle
        $u^*$ are smaller than 0.01.}
\begin{tabular}{llcccccc}
\hline\hline
     Phase 1  & Phase 2  & $p^*$ & $T^*$ & $\rho^*_1$ & $u_1^*$ & $\rho^*_2$ & $u_2^*$ \\
\hline
     fluid & bcc &  2.01 &  0.20 & 0.959 & -0.30 &  0.999 & -0.99 \\
     fluid & bcc &  1.00 &  0.18 & 0.811 & -0.32 &  0.922 & -1.28 \\
     fluid & bcc &  0.40 &  0.16 & 0.652 & -0.34 &  0.833 & -0.92 \\
     fluid & bcc &  0.05 &  0.13 & 0.333 & -0.35 &  0.868 & -1.51 \\
     fluid & bcc &  0.001 &  0.10 & 0.010 & -0.05 &  0.881 & -1.63 \\
\hline
     fluid & fcc-d &  4.60  &  0.40 &  0.972 & -0.07 & 1.072 & -0.13 \\
     fluid & fcc-d &  3.30  &  0.30 &  0.959 & -0.16 & 1.055 & -0.22 \\
     fluid & fcc-d &  1.99  &  0.20 &  0.936 & -0.31 & 1.034 & -0.38 \\
\hline
     bcc & fcc-o & 3.80  &  0.13 &  1.061 & -1.34 & 1.164 & -0.92 \\
     bcc & fcc-d & 3.60  &  0.15 &  1.054 & -1.28 & 1.153 & -0.85 \\
     bcc & fcc-d & 2.80  &  0.18 &  1.022 & -1.18 & 1.118 & -0.42 \\
     bcc & fcc-d & 2.00  &  0.20 &  0.981 & -1.08 & 1.030 & -0.38 \\
 \hline\hline  
\end{tabular}  
\end{table}

	Starting from the coexistence points given in Table \ref{tbl_CP}
the whole coexistence lines were obtained using the Gibbs-Duhem integration
method. Some coexistence points calculated by this method are given 
in Table \ref{tbl_GD}. As a test,
the melting point of the solid phases that are in coexistence 
with the fluid (i.e.\ the bcc solid and the fcc plastic crystal) was
also computed using the direct coexistence method. The melting points
obtained by this route are shown in Table \ref{tbl_DC}.  
As can be seen, free energy calculations and the direct coexistence method give
results that are consistent within statistical uncertainty.

\begin{figure}[t]
\begin{center}
\includegraphics[width=85mm,angle=0]{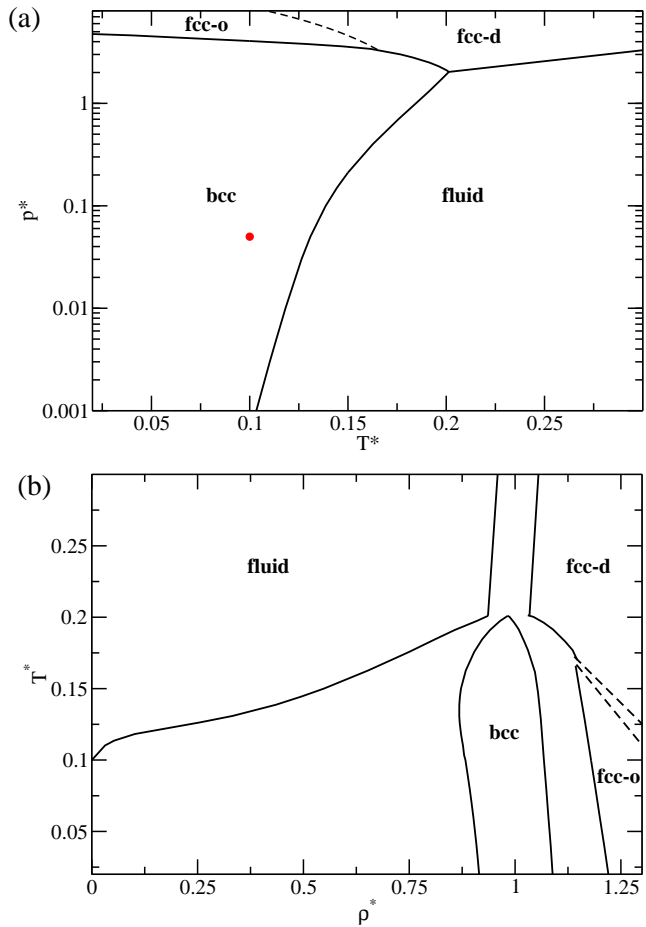}
\caption{\label{PD_0.3} Phase diagram of model patchy particles
	with tetrahedral symmetry for the 12-6 model and 
	a patch width $\sigma_{\rm pw}$=0.3 radians
        as a function of (a) pressure and temperature, and (b) temperature and
        density. The red dot in (a) indicates the thermodynamic state point at
	which direct coexistence simulations of the fluid-diamond and
	fluid-bcc interface were performed to study the growth
	behaviour of these solids.}
\end{center}
\end{figure}

\begin{table}[!tb]
\centering
\scriptsize
\caption{\label{tbl_DC} Coexistence points for the 12-6 model obtained using the
        direct coexistence method. For comparison, the results
        from free energy calculations are also given.}
\begin{tabular}{llcccccc}
\hline\hline 
        & & \multicolumn{2}{c} {Direct coexistence} &  & &
\multicolumn{2}{c} {Free energy calculations} \\ 
\hline
     Phase 1  & Phase 2  & $T^*$ & $p^*$ & & & $T^*$ & $p^*$ \\
\hline
     fluid &   bcc   & 0.14(1)&  0.21 &  & & 0.15  & 0.21(2)\\
     fluid &   fcc-d & 0.50   &  6.0(1) &  & & 0.50  & 5.91(2) \\
 \hline\hline
\end{tabular}
\end{table}

The complete phase diagram for the tetrahedral model with a
patch width $\sigma_{\rm pw}$=0.3 radians is shown in Fig.\ \ref{PD_0.3}.
All the solids considered, except diamond, are
stable over a region of the phase diagram. At low temperatures, the fluid
freezes into the bcc crystal, which upon compression is destabilized with
respect to the fcc-o solid. The fcc-o structure transforms into a plastic
crystal (fcc-d) at approximately $T^*$=0.16.
This order-disorder transition
is a first order transition; it exhibits a discontinuity both in
the energy and in the density. 
As this region of the phase diagram is not the focus of this work,
the temperature at which the order-disorder transition occurred
at a given pressure was estimated simply as the midpoint of the hysteresis
loops in the variation of the energy and density with temperature 
upon heating and quenching. 
The phase diagram exhibits a triple
point at $T^*$=0.201 and $p^*$=2.03, at which the fluid, the bcc and the 
fcc-d solids coexist. Above this temperature, the fluid freezes into the 
fcc-d plastic crystal.

The phase diagram of the tetrahedral particles is simplified with
respect to that of the six-patch octahedral particles that
we computed previously.\cite{JCP_127_054501_2007} 
Additional features for the octahedral system include
the stabilization of a low density crystal (a simple-cubic solid) and
reentrant behaviour for the coexistence lines between the fluid and simple-cubic
solid, and between the bcc and fcc crystals.
Another difference between the two models is that the bcc solid
is almost incompressible at zero temperature for the octahedral model,
whereas for the tetrahedral model the bcc can be compressed from $\rho^*=$0.915
up to $\rho^*$=1.225, which correspond to nearest-neighbour
distances of 1.124$\,\sigma_{\rm LJ}$ and 1.061$\,\sigma_{\rm LJ}$, respectively.
This means that the bcc solid can be compressed considerably from
the minimum energy structure (i.e.\ that for which nearest neighbours are
located at the distance of the minimum of the potential 1.123$\,\sigma_{\rm LJ}$).
The different behaviour 
is because the interpenetration
of the two diamond sublattices does not have an energy penalty
but the interpenetration of two simple cubic sublattices does.

However, there are also strong similarities between the two phase diagrams
and quantitative comparisons can be made. 
For example, the bcc phase is stable up to $T^*$=0.336
for the octahedral particles, whereas for the tetrahedral
particles it is stable up to $T^*$=0.201. As tetrahedral
particles have only four patches whereas octahedral have six patches,
one would expect that the maximum temperature 
for which the bcc solid is stable for the tetrahedral
particles should be roughly two thirds of that for octahedral
particles, which is in agreement with our results.
Note that in both cases the bcc solid is stabilized
by its low internal energy achieved 
because the patches can directly point at the neighbouring particles.

As already noted, the phase diagram obtained
here is somewhat different from the phase diagram 
recently reported by Romano \emph{et al.} for similar patchy tetrahedral 
particles.\cite{JPCB_113_15133_2009}
In contrast to our results, these authors found that the diamond crystal 
is stable over a region of the phase diagram. 
What is the origin of these differences? 
Romano \emph{et al.} used a different model potential from the
one studied here. In particular, they considered the Kern-Frenkel model,
in which particles are described as hard-spheres with some attractive
patches at the surface modeled by a square-well potential in both
relative orientation and interparticle distance.\cite{JCP_118_9882_2003}
Taking a fixed value for the patch width of 0.4 radians, these 
authors calculated the phase diagram for different ranges of
the potential (from 0.03\,$\sigma_{\rm HS}$ to 0.24\,$\sigma_{\rm HS}$, where
$\sigma_{\rm HS}$ is the diameter of the hard-spheres), and 
found that the bcc solid phase is destabilized with respect to
the diamond structure and the fluid phase as the range of the 
potential decreases.\cite{JPCM_19_322101_2007} 
Therefore, this suggests that the diamond structure might become
stabilized in our model if the range of the potential was decreased.

\begin{figure}[!hbt]
\begin{center}
\includegraphics[width=85mm,angle=0]{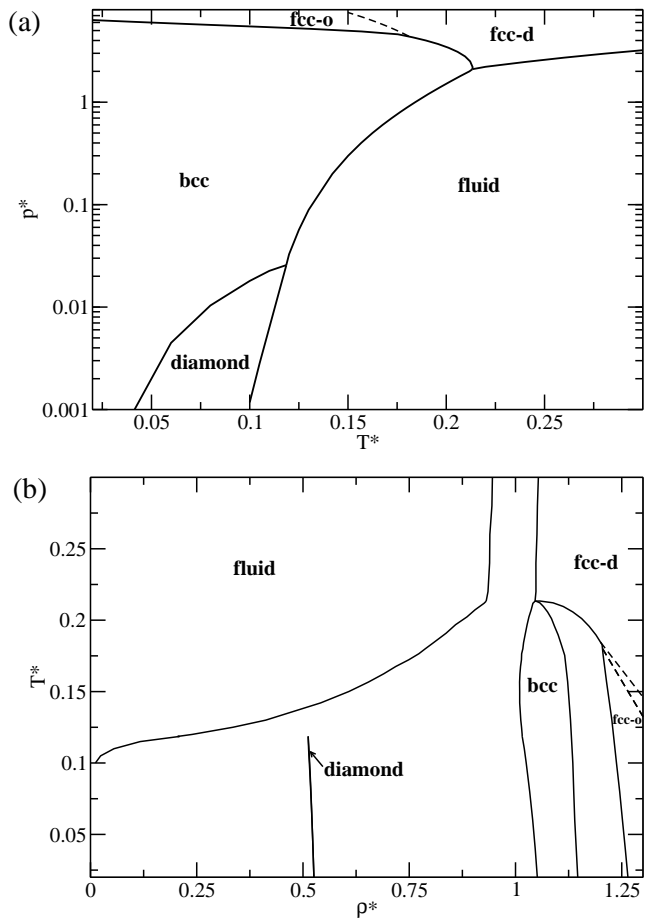}
\caption{\label{PD_20_10} Phase diagram of model patchy particles
	with tetrahedral symmetry for the 20-10 model and 
	a patch width $\sigma_{\rm pw}$=0.3 radians
        as a function of (a) pressure and temperature, and (b) temperature and
        density.}
\end{center}
\end{figure}

\begin{figure}[!h]
\begin{center}
\includegraphics[width=85mm,angle=0]{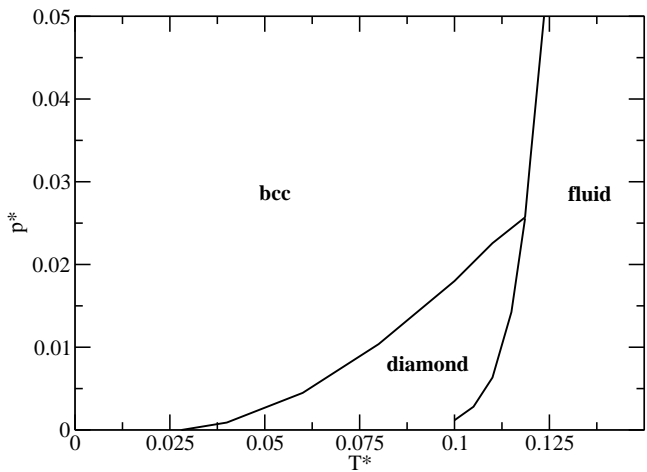}
\caption{\label{PD_enlarged} Enlarged view of the low temperature and
	low pressure region of the phase diagram for the
	20-10 model.}
\end{center}
\end{figure}

\begin{table}[t]
\centering
\caption{\label{tbl_free_energies2} Helmholtz free energies ($A_{sol}$) 
of the solid phases, as obtained by the Einstein molecule method,
for the 20-10 model.
The coupling parameters in the Einstein
molecule were chosen as $\lambda_t/(k_BT/\sigma_{\rm LJ}^2)=\lambda_o/(k_BT)$=70000.
The free energy $A_{sol}$  and the 
average potential energy ($U$) are given in units of $Nk_BT$. 
(TI) refers to independent simulations run as consistency checks.
Uncertainties in the free energies and potential energies are of
the order of 0.02 $NkT$.}
\begin{tabular}{lcccccccccc}
\hline\hline
Structure & & $T^*$  & & $p^*$ & & $\rho^* $ & & $U$ & &  A$_{sol}$ \\
\hline
  diamond      & & 0.10  & & 0.473  & & 0.530  & &  -16.57  & & -4.35 \\
  diamond      & & 0.10  & & 0.01   & & 0.515  & &  -16.47  & & -4.47 \\
  diamond (TI) & & 0.10  & & 0.01   & & 0.515  & &          & & -4.47 \\
  diamond      & & 0.13  & & 0.172  & & 0.521  & &  -11.65  & & -0.76 \\
\hline
  bcc          & & 0.10  & & 0.466 & & 1.040  & &  -16.86  & & -4.26 \\
  bcc          & & 0.10  & & 1.300 & & 1.065  & &  -16.93  & & -4.10 \\
  bcc (TI)     & & 0.10  & & 1.300 & & 1.065  & &          & & -4.10 \\
  bcc          & & 0.13  & & 0.284 & & 1.021  & &  -12.02  & & -0.58 \\
\hline
  fcc-o        & & 0.10  & & 5.234 & & 1.234  & &  -10.83  & & 1.78 \\
  fcc-o        & & 0.10  & & 8.000 & & 1.272  & &  -9.61   & & 3.35 \\
  fcc-o (TI)   & & 0.10  & & 8.000 & & 1.272  & &          & & 3.35 \\
  fcc-o        & & 0.13  & & 5.675 & & 1.234  & &  -7.59   & & 4.18 \\
\hline
  fcc-d        & & 0.50  & & 5.423 & & 1.050  & &  -0.14   & &  4.46 \\
  fcc-d        & & 0.50  & & 8.500 & & 1.163  & &  -0.09   & &  5.70 \\
  fcc-d (TI)   & & 0.50  & & 8.500 & & 1.163  & &          & &  5.71 \\
 \hline\hline
\end{tabular} 
\end{table}

\begin{table}[!ht]
\scriptsize
\centering
\caption{\label{tbl_CP2} Coexistence points for the 20-10 model
	obtained using 
        thermodynamic integration together with the Helmholtz free 
        energies given in Table \ref{tbl_free_energies2}.
	Uncertainties in the densities
	$\rho^*$ are of the order of 0.001 and in the potential energy per particle
	$u^*$ are smaller than 0.01.}
\begin{tabular}{llccccccccccccc}
\hline\hline
     Phase 1  & & Phase 2 & & $T^*$ & & $p^*$ & & $\rho^*_1$ & & $u^*_1$ & & $\rho^*_2$ & & $u^*_2$ \\
\hline
     fluid   & &  diamond & & 0.10 & &  0.001(1)  & & 0.010  & & -0.04 & & 0.515 & & -1.65 \\
     diamond & &  bcc     & & 0.10 & &  0.02(2)  & & 0.516  & & -1.65 & & 1.025 & & -1.67 \\
     bcc     & &  fcc-o   & & 0.10 & &  5.49(6)   & & 1.131  & & -1.57 & & 1.238 & & -1.07  \\
\hline
     fluid   & &  bcc     & & 0.13 & &  0.09(1)  & & 0.404  & & -0.31  & & 1.013 & & -1.55 \\
     bcc     & &  fcc-o   & & 0.13 & &  5.22(6)   & & 1.126  & & -1.48 & & 1.227 & & -1.00 \\
     fluid   & &  fcc-d   & & 0.50 & &  5.70(6)   & & 0.962  & & -0.02 & & 1.065 & & -0.07 \\
 \hline\hline
\end{tabular}
\end{table}

\begin{table}[ht]
\centering
\caption{\label{tbl_GD2} Coexistence points for the 20-10 model
	obtained using the
        Gibbs-Duhem method. Uncertainties in the densities $\rho^*$ are of the order of
        0.001 and in the potential energy per particle
        $u^*$ are smaller than 0.01.}
\begin{tabular}{llcccccc}
\hline\hline
     Phase 1  & Phase 2  & $p^*$ & $T^*$ & $\rho^*_1$ & $u_1^*$ & $\rho^*_2$ & $u_2^*$ \\
\hline
     fluid & bcc &  1.0 &  0.184 & 0.806 & -0.258 &  1.019 & -1.334 \\
     fluid & bcc &  1.4 &  0.197 & 0.837 & -0.246 &  1.023 & -1.274 \\
     fluid & bcc &  1.6 &  0.202 & 0.886 & -0.240 &  1.032 & -1.248 \\
\hline
     diamond & bcc &  0.023 &  0.110 &  0.514 & -1.604 & 1.021 & -1.633 \\
     diamond & bcc &  0.010 &  0.080 &  0.519 & -1.730 & 1.033 & -1.746 \\
     diamond & bcc &  0.001 &  0.040 &  0.524 & -1.874 & 1.046 & -1.877 \\
\hline
     bcc & fcc-o & 6.45  &  0.010 &  1.148 & -1.762 & 1.269 & -1.219 \\
     bcc & fcc-d & 5.88  &  0.060 &  1.139 & -1.658 & 1.252 & -1.146 \\
     bcc & fcc-d & 5.20  &  0.130 &  1.129 & -1.457 & 1.227 & -1.004 \\
     bcc & fcc-d & 4.00  &  0.190 &  1.100 & -1.281 & 1.189 & -0.410 \\
     bcc & fcc-d & 2.50  &  0.212 &  1.057 & -1.187 & 1.088 & -0.325 \\
\hline
     fluid & fcc-d & 2.59  &  0.250 &  0.937 & -0.043 & 1.051 & -0.095 \\
     fluid & fcc-d & 3.84  &  0.350 &  0.952 & -0.104 & 1.056 & -0.155 \\
     fluid & fcc-d & 5.08  &  0.450 &  0.955 & -0.182 & 1.060 & -0.247 \\
 \hline\hline  
\end{tabular}  
\end{table}

We checked this hypothesis by also calculating 
the phase diagram for a shorter-ranged model, where the LJ
model was replaced by a generalized LJ potential with exponents
of 20 and 10, rather than 12 and 6. 
The free energies at some selected thermodynamic states are given 
in Table \ref{tbl_free_energies2}. Coexistence points calculated from these data are shown
in Table \ref{tbl_CP2}, and coexistence lines obtained from Gibbs-Duhem
simulations are given in Table \ref{tbl_GD2}. The melting point
of the solid phases in coexistence with the fluid was checked by also performing
direct coexistence simulations. The agreement between the two routes
was satisfactory (see Table \ref{tbl_DC2}). The complete phase
diagram is shown in Fig.\ \ref{PD_20_10}. Besides
the bcc and fcc solids found for the 12-6 model, the diamond crystal
is now stable over a region of the phase diagram. For sufficiently
low pressures, the diamond structure is stabilized over the bcc at finite
temperatures. As can be seen in the temperature and density phase diagram, 
the diamond structure is only stable for a very narrow range of densities,
which is a consequence of its low compressibility.
An enlarged view of the pressure and temperature phase diagram (see Fig.\ \ref{PD_enlarged})
shows that the coexistence between the diamond and bcc solid phases occurs
at negative pressures for temperatures below about $T^*=$0.03, which means
that the bcc is the solid stable phase at very low temperatures.
As mentioned in Sec.\ \ref{subsect:solids}, at zero temperature
and pressure
the bcc phase is more stable than diamond and the same occurs at positive 
pressures because the former has a lower energy.

As for the bcc solid, its region of coexistence moves to
higher densities  because the minimum of the potential moves to shorter 
distances.  At temperatures close to zero, the bcc solid is stable 
for densities between $\rho^*$=1.051 and $\rho^*$=1.146,
which correspond to nearest-neighbour distances of 1.072$\,\sigma_{\rm LJ}$
(the distance of the minimum of the 20-10 LJ model)
and 1.043$\,\sigma_{\rm LJ}$, respectively.  As before, the bcc solid
can compress considerably before losing its stability with
respect to the fcc solid. 
It can also be seen that the bcc solid gains some
stability with respect to the fcc solid as the range of the potential
shortens (i.e., the bcc-fcc phase transition moves to higher pressures
as the range decreases). Again the fcc solid exhibits
an order-disorder transition at $T^*\approx$0.17.
The thermodynamic states of the two triple points calculated for
the 20-10 model are given in Table \ref{tbl_TP2}.
  
\begin{table}[tb]
\scriptsize
\centering
\caption{\label{tbl_DC2} Coexistence points for the 20-10 model
        obtained using the
        direct coexistence method. For comparison, the results
        from free energy calculations are also given.}
\begin{tabular}{llcccccc}
\hline\hline 
        & & \multicolumn{2}{c} {Direct coexistence} &  & &
\multicolumn{2}{c} {Free energy calculations} \\ 
\hline
     Phase 1  & Phase 2  & $T^*$ & $p^*$ & & & $T^*$ & $p^*$ \\
\hline
     fluid &   bcc   & 0.15   &  0.30(1) &  & & 0.15  & 0.30 \\
     fluid &   fcc-d & 0.30   &  3.2(1) &  & & 0.30  & 3.22 \\
 \hline\hline
\end{tabular}
\end{table}

\begin{table}[!tb]
\centering
\caption{\label{tbl_TP2} Thermodynamic properties of the triple points
	found for the 20-10 model.}
\begin{tabular}{lllccccc}
\hline\hline 
     Phase 1  & Phase 2  & Phase 3 & $T^*$ & $p^*$ & $\rho_1^*$ & $\rho_2^*$ & $\rho_3^*$ \\
\hline
     fluid &   diamond & bcc    &  0.119  & 0.026 & 0.205  & 0.512 & 1.016 \\
     fluid &   bcc     & fcc    &  0.213  & 2.10  & 0.931  & 1.046 & 1.046 \\
 \hline\hline
\end{tabular}
\end{table}

\begin{figure}[!tb]
\begin{center}
\includegraphics[width=85mm,angle=0]{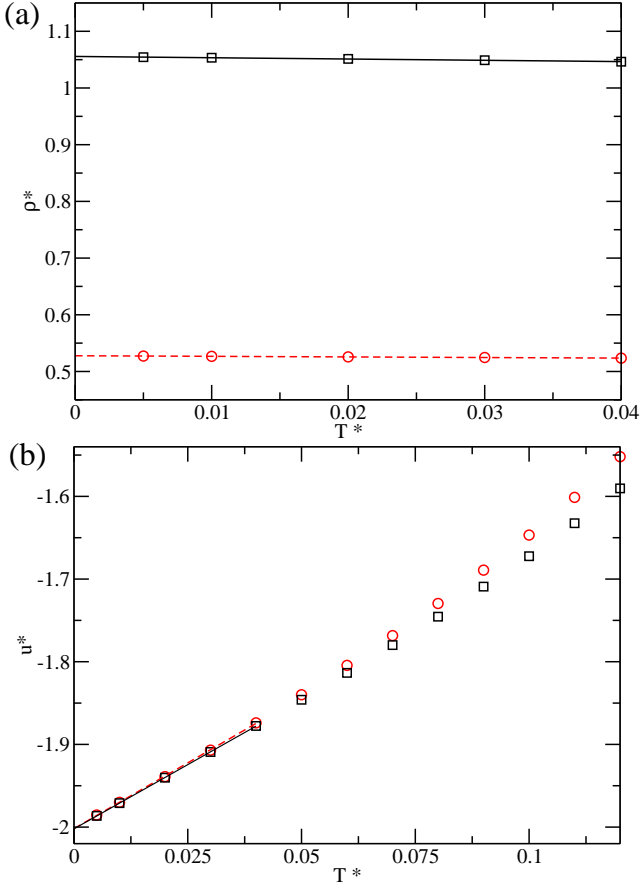}
\caption{\label{whalley} (a) Densities and (b) energies of the bcc
solid (squares and solid line) and of the diamond solid (circles and
dashed line) along the isobar $p^*=$0 for the 20-10 
model.}
\end{center}
\end{figure}

We have seen that,
as for the KF
model, the diamond structure is stabilized with respect to the
bcc solid when the range of the interactions decreases. 
But why is this so? 
A quick route to obtain information about the phase diagram of
a given model is by calculating the properties of the competing solid
phases at zero temperature.\cite{JCP_127_154518_2007}
At zero temperature,  the condition of chemical equilibrium is 
given by the equality of enthalpy of the phases in coexistence:
\begin{eqnarray}
& & U_I (p_{eq}, T=0) + p_{eq} V_I (p_{eq}, T=0) = \nonumber \\
& & U_{II} (p_{eq}, T=0) + p_{eq} V_{II} (p_{eq}, T=0) 
\end{eqnarray}
Therefore, phase transitions can be located at zero temperature
without computing free energies, just by calculating
the density and potential energy of both phases in coexistence.
Assuming that the change of internal energy and the change
of volume between the
two phases are independent of pressure, the coexistence pressure 
at zero temperature can 
be approximately calculated using:\cite{JCP_84_4087_1984}
\begin{equation}
p_{eq} = -\frac{\Delta U (p=0, T=0)}{\Delta V (p=0, T=0)}
\end{equation}
Using this expression
we have calculated the coexistence pressure at zero
temperature between the diamond
and bcc solids for the models studied in this work.
The densities and energies at zero temperature and pressure were estimated
by performing simulations at temperatures between $T^*$=0.04 and
$T^*$=0.005 and linearly extrapolating these data
to $T^*$=0 (see Fig.\ \ref{whalley}). The densities and internal
energies obtained using this procedure are given in Table \ref{tbl_T0}. 
It is found that, as expected (see Section \ref{subsect:solids}), 
for both models the diamond-bcc
transition occurs at negative pressures. 
The coexistence pressure is about $p^*=$-0.009 for the 12-6 model
and about $p^*=$-0.0003 for the 20-10 model. 
The less negative coexistence pressure for the shorter-ranged potential
is a consequence of the significantly smaller difference in energy between the 
two crystals for the 20-10 model.
This indicates that, even at zero temperature,
the diamond solid is stabilized by decreasing the range of the interactions.
It also provides a possible recipe to predict the stability of
the diamond solid: the less negative the coexistence pressure, the
higher the probability of stabilizing the diamond solid.
It is interesting to note that the behavior of our model and
the KF model at zero temperature will be different. In the KF model,
the diamond and the bcc crystals exhibit the same energy at zero temperature,
and so the phase transition occurs exactly at zero pressure, 
i.e.\ for any positive pressure the bcc solid will be more stable than the 
diamond, which will become more stable than the bcc only at negative pressures.

\begin{table}[!tb]
\centering
\caption{\label{tbl_T0} Density and potential energy at zero temperature
	and pressure for the diamond and bcc solids for the two studied model potentials.}
\begin{tabular}{lllcccc}
\hline\hline 
     Model &  & Solid  & & $\rho^*$ & & $u^*$ \\
\hline
     12-6  &  & diamond  & & 0.4599  & & -1.9765 \\
           &  & bcc      & & 0.9206  & & -1.9867 \\
\hline
     20-10 &  & diamond  & & 0.5277  & & -2.0019 \\
           &  & bcc      & & 1.0557  & & -2.0021 \\
 \hline\hline
\end{tabular}
\end{table}

\begin{figure}[!tb]
\begin{center}
\includegraphics[width=85mm,angle=0]{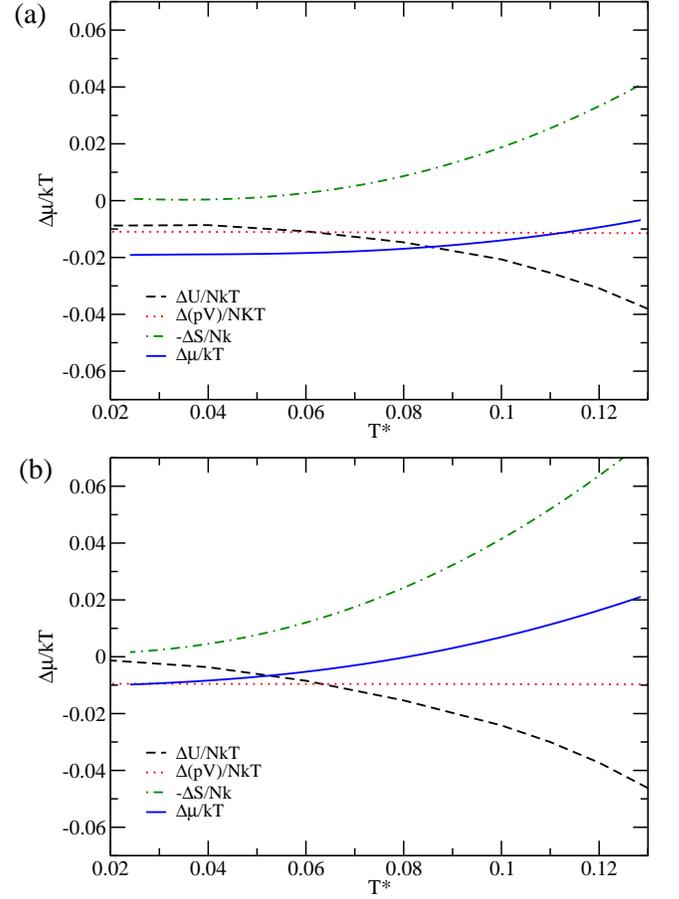}
\caption{\label{contributions} 
        Chemical potential difference between the bcc and diamond solids
	along the $p^*$=0.01 isobar for (a) the 12-6 
	model and (b) the 20-10 
	model. The contributions of potential energy, $pV$ and entropy to
	the chemical potential are also given.}
\end{center}
\end{figure}

To further understand the origin of the stabilization of
the diamond structure at finite temperature as the range of the interactions decreases,
we have calculated the different contributions to the
chemical potential for both the bcc and diamond solids along the 
$p^*=$0.01 isobar, a pressure at which diamond is stable over some 
temperature range for the 20-10 model. The difference between the chemical
potential of the bcc and the diamond solids 
($\Delta \mu= \mu^{bcc}-\mu^{diamond}$) and its three contributions
($\Delta \mu /kT = \Delta U/NkT + \Delta (pV)/NkT - \Delta S/Nk$) are 
shown in Fig.\ \ref{contributions}. The chemical
potential is computed by thermodynamic integration along the isotherm $p^*=$0.01
starting from the free energy at $T^*$=0.10.

Results for both the long-ranged 12-6 LJ model and the shorter-ranged 20-10 
model are shown. At this pressure, $p^*$=0.01, $\Delta \mu/kT$
is negative over all the temperature range for the 12-6 model
(i.e.\ the bcc solid is more stable than diamond),
and passes from negative to positive at $T^*\approx 0.08$ for the
20-10 model (i.e.\ for temperatures below $T^*\approx 0.08$
the bcc solid is most stable but for temperatures above this the diamond 
structure is most stable). 
Analyzing the contributions to the chemical potential,
it can be seen that the $\Delta (pV)/NkT$ term is practically independent of the temperature (for the
range of temperatures considered) and, at this pressure,
its value is very similar for the
two models studied. The important term for understanding the thermal 
stabilization of the diamond structure is $\Delta S$. In the harmonic
approximation, the vibrational entropy is given by 
\begin{equation}
S_{\rm vib}=(6N-3)k\left(\ln\left(\frac{kT}{h\bar\nu}\right) +1 \right)
\end{equation}
where $\bar\nu$ is the geometric mean vibrational frequency.
Note that in a classical statistical mechanics formalism (as is appropriate 
for the simulations considered in this work), 
the entropy tends to minus infinity at zero temperature. 
However, the difference of entropy between two phases can remain finite. 
In particular, it follows that, again in the harmonic approximation, 
\begin{equation}
\Delta S_{\rm vib}=(6N-3)k\ln\left(\frac{\bar\nu_{\rm diamond}}
{\bar\nu_{\rm bcc}}\right) 
\label{eq:dSvib}
\end{equation}
As Fig.\ \ref{contributions} shows, $\Delta S$ approaches zero as the 
temperature is decreased. The vibrational frequencies are
related to the curvature of the potential energy surface at the minimum
corresponding to the crystal. At this minimum the two sublattices of
the bcc crystal interact only very weakly, and so the vibrational frequencies 
are essentially the same for the two crystals, except for the three modes 
corresponding to the displacement of the two sublattices with respect to each 
other in the bcc crystal. Hence, $\Delta S(T=0)\approx 0$.

Eq.\ (\ref{eq:dSvib}) also implies that $\Delta S_{\rm vib}$ will not vary with
temperature if the vibrations are harmonic. However, as is clear from 
Fig.\ \ref{whalley}, the variation of the potential energy with temperature
deviates from the linearity expected for harmonic vibrations, and
it is noticeable that this deviation is more pronounced for the diamond 
lattice, i.e.\ the vibrations in the diamond lattice are more anharmonic.
As the magnitude of the vibrations increase with temperature, the presence
of another sublattice becomes an increasing constraint on the thermal motion
in the bcc crystal. By contrast, the empty space in the diamond lattice 
allows the magnitude of the vibrations to increase more rapidly than that
for a harmonic system. It is this greater vibrational entropy that drives
the thermal stabilization of the diamond phase. 
There is one more subtlety: the increase of $-\Delta S$ with temperature is 
partially offset by a corresponding decrease in $\Delta U$, because the greater
vibrational entropy available as the energy increases causes the energy to
increase more rapidly than for harmonic vibrations. This relationship between
$\Delta S$ and $\Delta U$ is of course inherent in the formulae: 
$(\partial S/\partial T)_P=C_p/T$ and 
$(\partial U/\partial T)_P=C_p$.

This analysis leaves one question: why is this thermal stabilization 
of the diamond crystal more pronounced for the shorter-range potential. 
Firstly, the zero-temperature difference in potential energy between
the two crystals is reduced. Secondly, as
we noted earlier the density difference between the bcc crystal 
when at its potential energy minimum and when the repulsive cores start to 
overlap is smaller for the 20-10 model. Hence, the magnitude of the vibrations
required for the two sub-lattices to begin to interact significantly is smaller.
The potential effect of this difference is reduced because the magnitude of 
the bond-stretching vibrations is also reduced for the 20-10
model because the potential is stiffer as a function of distance. 
However, the patch width is the same for the two models and hence the 
magnitude of the angular vibrations will be similar for the two models,
and it is the effect of the reduced ``room'' on the angular vibrations in the
bcc crystal that leads to the enhanced entropic stabilization of the diamond 
structure for the shorter-ranged model.
 
In summary, our results show significant similarities 
with those of Romano \emph{et al.}\cite{JPCB_113_15133_2009} 
Firstly, we find that at low pressure, the bcc and diamond crystals
have very similar free energies.
Secondly, we found that the diamond solid is stabilized as the range of
the potential decreases. 
However, the main difference with their results is that the region of diamond 
stability is significantly reduced, and that for 
a sufficiently long-ranged potential, the diamond crystal is never 
thermodynamically the most stable phase.
This difference is most likely due to the different shapes of the potential wells
for the two models.

In the KF model, the patches are modeled as square wells, and
the flat-bottomed nature of these wells means that, if no bonds are broken
(in the work of Vega and Monson using the PMW model they indeed found that
in both the diamond and bcc solids breaking of bonds was a rare event\cite{vegamonson}),
the configuration space available to the solid at a given density is 
independent of temperature.
Thus, there is an entropy term associated with the rattling of the molecules 
in these square wells even at zero temperature. Furthermore, this entropy term
will favour the diamond crystal, because of the reduction in the configuration
space available to the bcc solid due to interactions between the two 
sub-lattices. 
By contrast, for our patchy LJ model, the system becomes localized in the
potential energy minimum corresponding to the respective solid at zero 
temperature, because all vibrations come with a potential energy cost.
Only as the temperature increases, and hence the amplitude of the vibrational
motion increases, does a difference in vibrational entropy favouring
diamond become apparent.

Now that the phase diagram for our patchy particles is known, 
it would also be interesting to
perform nucleation studies to investigate which solid structure nucleates
from the fluid at different thermodynamic states. This is
by no means a trivial question.
For example, a significant number of systems have been shown to follow 
Ostwald's step rule;\cite{ZPC_22_289_1897,ZPC_163_399_1933,OstwaldStepRule}
namely, that a fluid does not crystallize in the most thermodynamically stable phase if
there is an alternative that is separated from the fluid by a lower free energy 
barrier. Given the small free energy differences between the bcc and diamond
crystals at low pressure, it would not be so surprising if the selection of 
crystal form was dominated by kinetic effects in this region of the phase
diagram.

\begin{figure}[!tb]
\begin{center}
\includegraphics[width=65mm,angle=0]{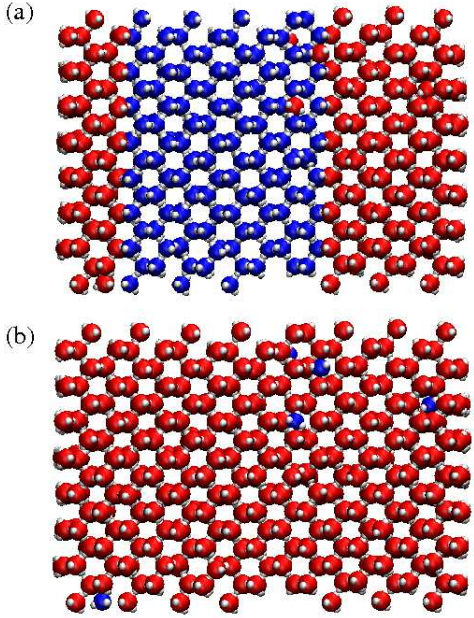}
\caption{\label{liq-diamond} Final configuration of the direct 
coexistence simulations of a fluid-diamond interface at $T^*$=0.1 and 
$p^*$=0.05 for the 12-6 model. Initially the simulation
box contained 512 molecules in a dimond solid (i.e., 4$\times$4$\times$4
unit cells) plus another 512 molecules in the fluid phase. Two different representations
are shown. In (a) molecules that were in the diamond crystal in
the initial configuration are coloured in blue, whereas those that
were fluid molecules in the initial configuration are shown in red.
In (b) molecules that belong to the same diamond sublattice are
coloured in red, whereas those molecules not connected to the sublattice 
(i.e., defects) are
coloured in blue. As can be seen, the diamond crystal grows with a small number of defects.}
\end{center}
\end{figure}

Although a full study of the kinetics of crystallization is beyond the scope
of this work, interesting information about the nature of crystal growth from 
the fluid can be obtained from additional direct coexistence simulations. 
Firstly, we examine whether a diamond crystal can continue
to grow with this structure, even though it is metastable with respect to the
bcc solid. The interface was simulated for the 12-6 model at $T^*$=0.1 
and $p^*$=0.05, a thermodynamic state at which the bcc solid is the most
stable phase (see the red dot in Fig.\ \ref{PD_0.3}(a)).
Fig.\ \ref{liq-diamond} clearly shows that the crystal growth
maintains the diamond lattice albeit with some defects. 
This result is in keeping with previous work,\cite{glotzer_langmuir2005} 
where it was shown that a diamond crystal could be grown by introducing
a crystalline seed with diamond structure into the simulation box. 

\begin{figure}[!h]
\begin{center}
\includegraphics[width=85mm,angle=0]{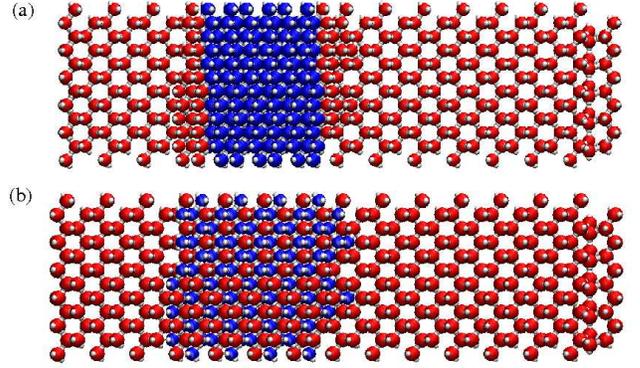}
\caption{\label{liq-bcc} Final configuration of the direct coexistence simulations 
of a fluid-bcc interface at $T^*$=0.1 and $p^*$=0.05 for a simulation box containing
1296 molecules for the 12-6 model. Initially the simulation box contained 432 molecules in a bcc
solid (i.e., 6$\times$6$\times$ 6 unit cells) and 864 molecules of fluid.
(a) and (b) show two different repesentations of the final configuration of
the simulation, in which all the fluid has crystallized.
(a) The molecules that were in the bcc solid structure in the initial
configuration are coloured in blue, whereas those that were fluid molecules
in the starting configuration are coloured in red. It can be seen that almost
all the fluid has solidified into a diamond crystal. Only one or two incomplete
bcc layers form at the two bcc-fluid interfaces. Most likely
the defects in these first layers make less and less probable the growth
of the bcc solid.
(b) As mentioned in the manuscript a bcc solid is formed by two interpenetrating
diamond solids. The two sublattices are highlighted by colouring the particles
belonging to each sublattice in a different colour, red for one sublattice and
blue for the other. It can be seen that in this particular example, the same sublattice
grew from each of the two interfaces. However, when the two diamond
crystals growing from the two interfaces meet, some defects appear because
as some particles were used to form one or two incomplete bcc layers at the bcc-fluid
interfaces, the number of available particles is incommensurate with the
dimensions of the simulation box (even though we chose it to be commensurate).}
\end{center}
\end{figure}

Secondly, we examine the nature of the crystal growth on a bcc crystal 
under the same low pressure conditions. This was motivated by our hunch that
the bcc/fluid interface might nucleate a diamond crystal that is coherent
with the bcc lattice. Our reasoning was that if a defect or fluctuation leads 
to one of the diamond sub-lattices of the bcc crystal outgrowing the other, it
may be hard to restore the bcc structure at the interface, because diffusion
of particles through the other sub-lattice would be very slow or even 
unfeasible. Instead, the `selected' sub-lattice would be more likely to continue
to grow, leading to a diamond crystal. The results of the direct coexistence
simulations of a bcc/fluid interface at low pressure reported in
Fig.\ \ref{liq-bcc} confirm this scenario. 

\begin{figure}[!h]
\begin{center}
\includegraphics[width=80mm,angle=0]{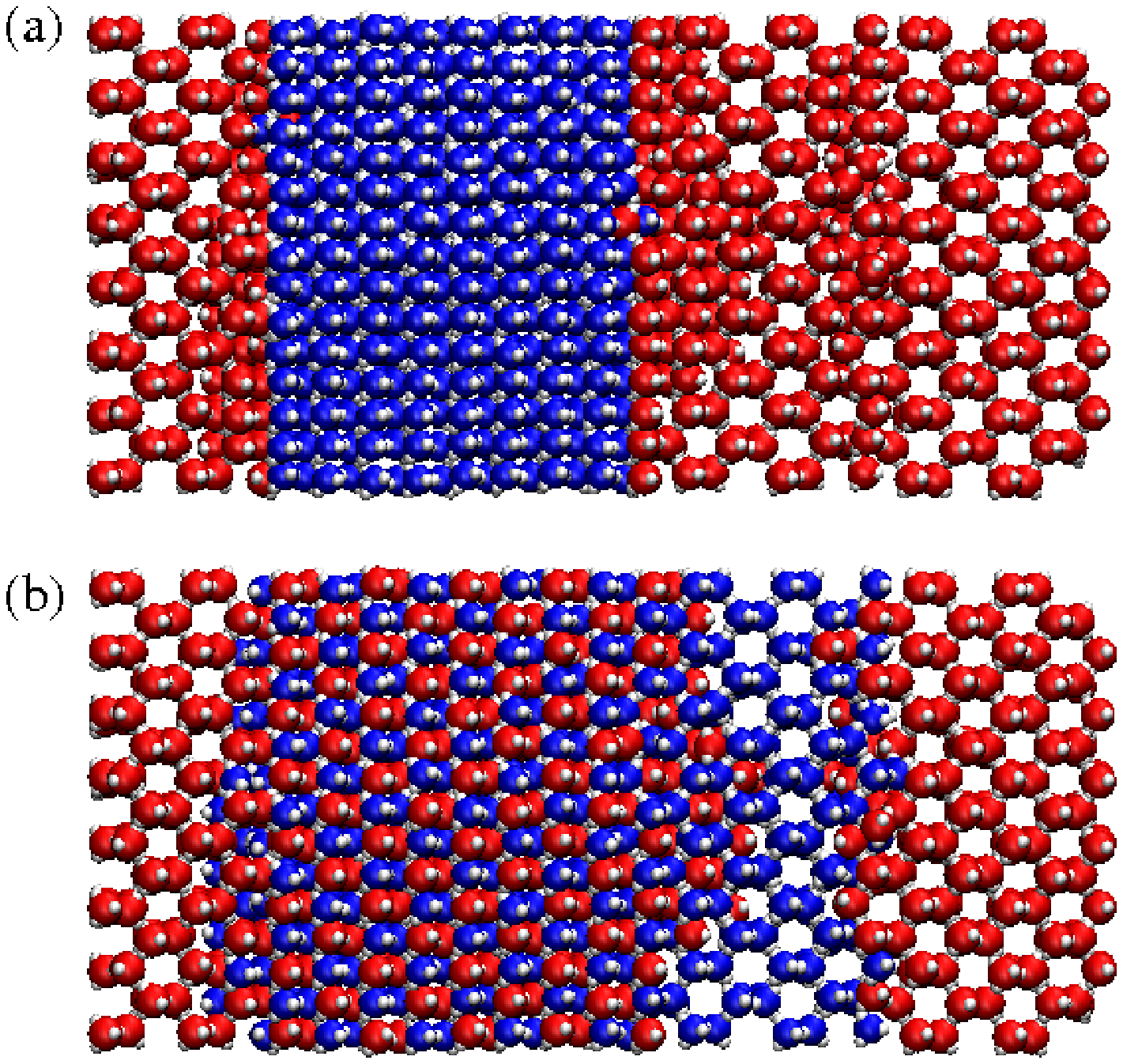}
\caption{\label{LJ126_8} Analysis of the bcc-diamond interface for the 12-6 model
for a simulation box containing
2048 molecules. Initially the simulation box contained 1024 molecules in a bcc
solid (i.e., 8$\times$8$\times$ 8 unit cells) and 1024 molecules of fluid.
As before, (a) and (b) show two different repesentations of the final configuration of
the simulation, in which all the fluid has crystallized.
(a) The molecules that were in the bcc solid structure in the initial
configuration are coloured in blue, whereas those that were fluid molecules
in the starting configuration are coloured in red. It can be seen that almost
all the fluid has solidified into a diamond crystal. Only one or two incomplete
bcc layers form at the two bcc-fluid interfaces.
(b) The two sublattices are highlighted by colouring the particles
belonging to each sublattice in a different colour, red for one sublattice and
blue for the other. In contrast to the example discussed in the manuscript, a different
sublattice has grown from each of the bcc-fluid interfaces. Some
defects appear when the two lattices meet. It can be observed that one of the
sublattices grew much more than the other.}
\end{center}
\end{figure}

\begin{figure}[ht]
\begin{center}
\includegraphics[width=85mm,angle=0]{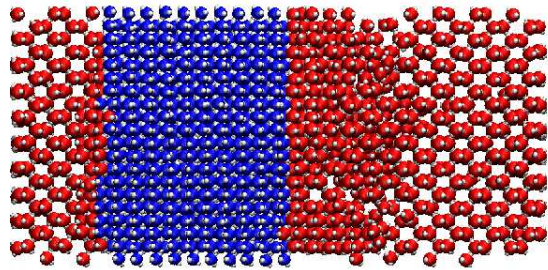}
\caption{\label{LJ126_10} Final configuration for the bcc-diamond interface for
the 12-6 model and a simulation box containing 4000 particles. Initially the simulation box
contained 2000 molecules in the bcc solid (i.e., 10$\times$10$\times$10 unit cells) 
and 2000 molecules in the fluid phase. 
Molecules that were in the bcc solid in the initial configuration are coloured
blue and those that were in the fluid phase in the initial configuration
are coloured in red.
As before, the fluid crystallizes 
in a diamond crystal with some defects.}
\end{center}
\end{figure}

\begin{figure}[!ht]
\begin{center}
\includegraphics[width=70mm,angle=0]{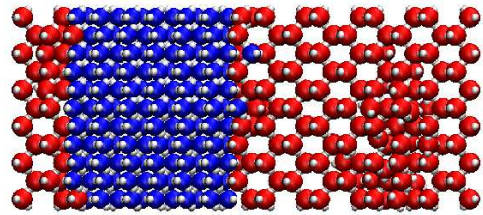}
\caption{\label{LJ2010_6} Final configuration for the bcc-diamond interface
for a simulation box containing 864 molecules for the shorted ranged LJ 20-10
model. Initially the simulation box contained 432 molecules in the bcc
solid (i.e., 6$\times$6$\times$ 6 unit cells) and other 432 molecules in the fluid phase. 
Molecules that were in the bcc solid in the initial configuration are coloured
blue and those that were in the fluid phase in the initial configuration
are coloured in red.}
\end{center}
\end{figure}

We checked that the growth of the diamond crystal from the bcc solid was not
an artifact caused by the small size of the system by performing
simulations for larger system sizes, which included both enlarging
the area of the interface and the distance between the two interfaces
in the simulation box. In all these examples the fluid in contact with the 
bcc solid crystallized in a diamond crystal (see Figs. \ref{LJ126_8} and
\ref{LJ126_10}).
In addition, we performed simulations for the shorter-ranged 
20-10 LJ model at $T^*$=0.10 and $p^*$=0.05 (for which bcc is the
most stable phase) and again it is observed that the bcc solid grows into a diamond
crystal (see Fig.\ref{LJ2010_6}). 

The growth of a bcc solid in contact with the fluid into a diamond crystal is
a stochastic process and, therefore, depending on the initial conditions,
the diamond crystal can grow following many possible different paths. Even 
though a detailed analysis of how the diamond crystal grows is beyond
the scope of this work, some useful information can be obtained by inspecting
the final configuration of our simulations. In all the simulations performed
we observe that one or two incomplete bcc layers grew on the two faces
of the bcc solid that were in contact with the fluid (snapshots are
provided in Figs.\ref{liq-bcc},\ref{LJ126_8},\ref{LJ126_10},\ref{LJ2010_6}). As discussed before, the vacancies that 
are left on these two layers are most likely responsible for the growth
of a diamond sublattice. It is observed that the diamond
crystal grows from both interfaces. In the example shown in Fig. \ref{liq-bcc}
the same diamond lattice grew from both interfaces. However, it is also
possible that a different sublattice grows from each interface and an example
is provided in Fig.\ref{LJ126_8}.
As there is not any reason why the same lattice should grow from the two
sublattices, there is a 50\%  probability  
that they would be the same and 50\% probability that they would be different.
It is also observed that the amount of growth from each interface 
is often different
and that when the crystals grown from the two
interfaces meet usually some defects appear irrespective of whether
the same or a different sublattice has grown from the two interfaces.

These results have important consequences for the formation of a diamond crystal 
from patchy tetrahedral particles, suggesting that the diamond phase may be able to form 
even when it is metastable with respect to other crystal structures.
For example, even if the low-pressure nucleation kinetics
were to favour the formation of bcc nuclei, these might well then grow into
diamond crystals. Furthermore, even if the low-pressure nucleation kinetics
were so slow that the system instead formed a glass (as perhaps suggested
by our previous annealing simulations\cite{jon}), an alternative pathway 
might be to use a bcc crystal that was generated at higher pressure as a seed.

\section{Conclusions}
\label{conclusions}

	The phase diagram of model tetrahedral patchy particles was obtained
from free energy calculations. Even though the width of the patches was 
narrow enough for the low-pressure crystal form to be dominated by the
energetics of specific patch-patch interactions, our results
indicate that the diamond crystal is only thermodynamically
stable when the range of the potential is below a critical value. 
At low pressures and finite temperatures, the diamond is
competitive with a bcc solid,
which consists of two interpenetrating diamond
lattices. In a diamond lattice, there is enough empty space to 
interpenetrate another diamond lattice without repulsive energy
between the two sublattices, thus obtaining a bcc crystal. 
Therefore, both the diamond and the bcc
exhibit very similar energies, but the bcc is stabilized over the
diamond lattice because of its lower enthalpy (i.e., lower
value of the $pV$ term). Only at finite temperatures
can the higher entropy of the diamond crystal make it more stable
than the bcc solid. Our results show that the difference of entropy between
the bcc and diamond solids increases as the range of the interactions decreases.
As a consequence the diamond solid is only stabilized when the range of the
interactions is below some given value.

	For the short-ranged model, the diamond solid is only stable
at low pressures and finite temperatures. On compression the diamond
transforms into the bcc solid.
The rest of the phase diagram is qualitatively similar for the
two ranges studied, although there are some quantitative differences.
It is found that, upon compression,
the bcc crystal transforms into an ordered fcc crystal, which exhibits
a somewhat higher energy because in this case the four
patches cannot be simultaneously perfectly aligned to four
nearest neighbours. This transition moves to higher pressures
as the range of the potential decreases. At high temperatures the fluid freezes
into a fcc plastic crystal.

The structure of our phase diagrams show strong similarities to 
those computed by Romano \emph{et al.} for a similar tetrahedral patchy
model. In particular, for both models the diamond structure is stabilized
with respect to the bcc solid as the range of the potential is decreased.
However, the differences between the two models --- the region of stability
for the diamond structure is smaller for our model --- also highlight that, 
as well as the symmetry of the particles, the particular `shape' of the 
interparticle potential can also have a strong effect
on the phase behaviour and on the stabilization of the diamond structure.
What type of potential is likely to be representative of the patchy colloids
that experimental groups are seeking to produce is not yet clear --- it will 
depend on how the different surfaces of the patchy colloids are 
functionalized in order to generate selective attractions between the patches.

	Even though we found that simple anisotropic models can
stabilize the diamond solid, the diamond phase is only stable for a very 
narrow range of pressures. This behaviour is in contrast with many water
models for which ice Ic (diamond structure) is found to
be more stable than ice VII (bcc structure) over a wider region
of the phase diagram.\cite{vegaprl04,jpcm_2005_17_S3283}
These water models consist of a LJ at the oxygen site
plus two positive point charges on the hydrogens sites
and a negative charge whose location depends on the particular model.
The stabilization of the diamond structure is related to the penetrability
of the water model. The hydrogen bond distance is about 2.7{\AA }, whereas
the LJ $\sigma$ parameter is 3.15{\AA }. As a consequence the
interpenetration of a second diamond sublattice has a large energy
penalty in water. 

	If one wants to nucleate a particular crystal from the fluid phase,
both kinetic and thermodynamic effects must be considered.
For the long-ranged 12-6 LJ model we have found that a diamond 
crystal can be grown by introducing a crystalline seed (be it of diamond or 
bcc structure) into the simulation box. This is good news for those seeking to 
produce colloidal diamond using patchy colloids.

It would be interesting to perform nucleation studies on this
system to further understand the crystallization behaviour.
As seen in previous work, the nucleation of a diamond crystal is
likely to be a challenging problem, because, besides the bcc solid,
the local structure in the liquid frustrates the nucleation
of either the diamond or the bcc crystal.\cite{jon}
The phase diagram calculated in this work is a necessary precursor 
to such nucleation studies.

	After submitting this manuscript, Romano \emph{et al.}\cite{JCP_132_184501_2010}
published an extended version of their previous work on the calculation 
of the phase diagram of tetrahedral particles described with the KF model.\cite{JPCB_113_15133_2009} 
They also propose an explanation for the higher vibrational
entropy of the diamond crystal with respect to the bcc solid, 
which is in line with the discussion in the present manuscript.

\acknowledgements

This work was funded by grants FIS2007-66079-C02-01 of Direcci\'on 
General de Investigaci\'on, MODELICO-CM P2009/ESP-1691 of
Comunidad Aut\'onoma de Madrid and 200980I099 of CSIC. 
JPKD and AAL are grateful for financial support from the Royal Society.
CV thanks Francesco Sciortino and Flavio Romano for useful
discussions during CV's visit to the University of Rome and for
sending him Ref. \onlinecite{JCP_132_184501_2010} prior to publication.
EGN also acknowledges useful discussions with 
No\'e G. Almarza.

\end{document}